\documentclass{egpubl}
\usepackage{eg2026}

\ShortPresentation      % uncomment for (final) Short Conference Presentation
\usepackage{amsmath}

\usepackage[T1]{fontenc}
\usepackage{dfadobe}

\usepackage{cite}  % comment out for biblatex with backend=biber
\BibtexOrBiblatex
\electronicVersion
\PrintedOrElectronic
\ifpdf \usepackage[pdftex]{graphicx} \pdfcompresslevel=9
\else \usepackage[dvips]{graphicx} \fi

\usepackage{egweblnk} 
\usepackage{xcolor}
\usepackage{float}
\usepackage{placeins}

\title{Prompt-Driven Color Accessibility Evaluation in Diffusion-based Image Generation Models}

\author[X. Zhuang \& J. Echevarria \& K. Ak\c{s}it]
{\parbox{\textwidth}{\centering Xinyao Zhuang\thanks{Corresponding author}$^{1}$\orcid{0009-0001-5180-6093}
       and Jose Echevarria$^{2}$\orcid{0000-0001-6802-0911}
       and Kaan Ak\c{s}it$^{1}$\orcid{0000-0002-5934-5500}
       }
       \\
{\parbox{\textwidth}{\centering $^1$University College London, United Kingdom\\
        $^2$Adobe Research, USA
      }
}
}
\begin{document}

% uncomment for using teaser
% \teaser{
%  \includegraphics[width=0.9\linewidth]{eg_new}
%  \centering
%   \caption{New EG Logo}
% \label{fig:teaser}
%}

\maketitle
%-------------------------------------------------------------------------

\begin{abstract}
   % The ABSTRACT is to be in fully-justified italicized text, 
   % between two horizontal lines,
   % in one-column format, 
   % below the author and affiliation information. 
   % Use the word ``Abstract'' as the title, in 9-point Times, boldface type, 
   % left-aligned to the text, initially capitalized. 
   % The abstract is to be in 9-point, single-spaced type.
   % The abstract may be up to 3 inches (7.62 cm) long. \\
   % Leave one blank line after the abstract, 
   % then add the subject categories according to the ACM Classification Index 
Generative models are increasingly integrated into creative workflows.
While text-to-image generation excels in visual quality and diversity, color accessibility for users with Color Vision Deficiencies (CVD) remains largely unexplored.
Our work systematically evaluates color accessibility in images generated by a common pretrained diffusion model, prompted to improve accessibility across diverse categories.
We quantify performance using established, off-the-shelf CVD simulation methods and introduce "CVDLoss", a new metric measuring differences in image gradients indicative of structural detail.
We validate CVDLoss against a commonly used daltonization method, demonstrating its sensitivity to color accessibility modifications.
Applying CVDLoss to model outputs reveals that existing diffusion models struggle to reliably respond to accessibility-focused prompts.
Consequently, our study establishes CVDLoss as a valuable evaluation tool for accessibility-aware image generation and post-processing, offering insights into current generative models' limitations in addressing color accessibility.
\ccsdesc[500]{Human-centered computing~Accessibility}
\ccsdesc[500]{Computing methodologies~Image processing}
\ccsdesc[500]{Computing methodologies~Machine learning}

\printccsdesc  

\end{abstract}  
%-------------------------------------------------------------------------

\section{Introduction}

% Diffusion-based generative image models have demonstrated impressive capabilities in producing visually compelling and diverse imagery from textual descriptions. Recent advances have enabled these models to generate content with high fidelity, complex structure, and rich color composition, leading to widespread adoption in creative, industrial, and scientific applications. However, despite their rapid progress, such models are primarily optimized for normal trichromatic vision and often overlook the accessibility needs of individuals with Color Vision Deficiencies (CVD). \jose{An alternate hopefully punchier version for this first paragraph could be along the lines: Existing image daltonization methods can successfully improve color accessibility. Diffusion-based models are impressive at capturing user intent, can they also do prompt-based daltonization? Thoughts?}

Existing image daltonization methods have demonstrated that post-processing color remapping can effectively improve visual accessibility for individuals with Color Vision Deficiencies (CVD) \cite{Ribeiro2019}. Meanwhile, diffusion-based generative image models have recently attained significant success in producing visually appealing, semantically rich, and highly colorful images directly from textual prompts. This naturally gives rise to a question: can generative models effectively perform accessibility-aware color transformations through prompt design alone, without requiring any explicit post-processing?

\begin{figure}
  \centering
  \includegraphics[width=\linewidth]{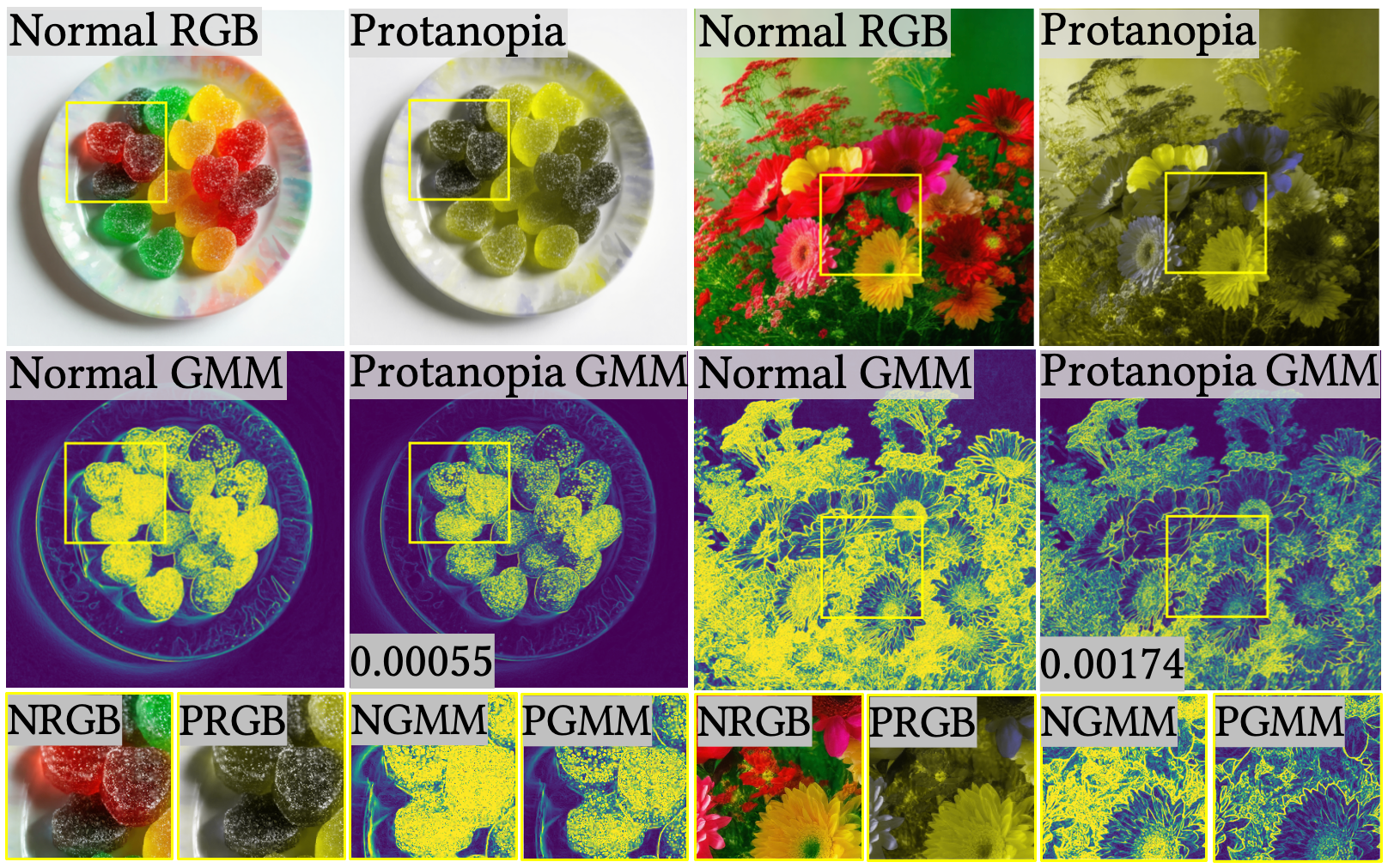}
  \caption{Qualitative illustration of CVDLoss on colorblind-aware generated images.
Two representative categories are shown under Normal vision 
% \jose{remove RGB. "RGB vision" is not a thing.} 
and Protanopia simulation, together with their corresponding Gradient Magnitude Maps (GMMs). The reported CVDLoss values in lower left corner quantify the discrepancy between normal and protanopia GMMs. Highlighted regions illustrate how local color gradients and edge structures alter for Protanopia. }
% \jose{Update with new metric. Choose image with lower metric.}
  \label{fig:2mse_example}
  \vspace{-4mm}
\end{figure}

CVD impact a considerable segment of the global population. The most common forms are protanopia (red-blindness) and deuteranopia (green-blindness)\cite{birch2012worldwide}. Individuals diagnosed with these conditions exhibit sensitivity to specific wavelength ranges, frequently resulting in challenges in differentiating colors that are typically separable under normal trichromatic vision. %\jose{The following sentences do not add much. Color accessibility is the same no matter how the image was generated.}Although accessibility considerations have been extensively studied for user interfaces, data visualization and static imagery, the accessibility of synthetically generated images, particularly those produced by modern diffusion models, remains relatively underexplored.
Most existing color accessibility guidelines focus on luminance contrast between foreground and background elements\cite{WCAG2018}. Despite their efficacy in numerous contexts, luminance-only criteria fail to capture perceptual conflicts arising from hue and saturation differences, which play a critical role for users with CVD. These conflicts are known to occur in complex visual scenes, including both natural images and generated imagery, where they may appear not only at object boundaries, but also within textures, fine details, and other visually salient regions.
% In generative imagery \jose{not only on generative imagery}, such conflicts may emerge not only at object boundaries but also within textures, fine details, and visually salient regions distributed across the image. 
As illustrated in Figure~\ref{fig:2mse_example}, diffusion-generated images can exhibit substantial perceptual changes under CVD simulation. The highlighted regions reveal that local textures and structural cues may be altered or weakened, motivating the need for an evaluation metric that captures any loss in perceptual structure.% beyond luminance alone. 
%Consequently, evaluating the accessibility of generated images necessitates metrics that extend beyond luminance and consider perceptual structure in a more comprehensive manner.

\begin{figure}[h]
  \centering
  \includegraphics[width=\linewidth]{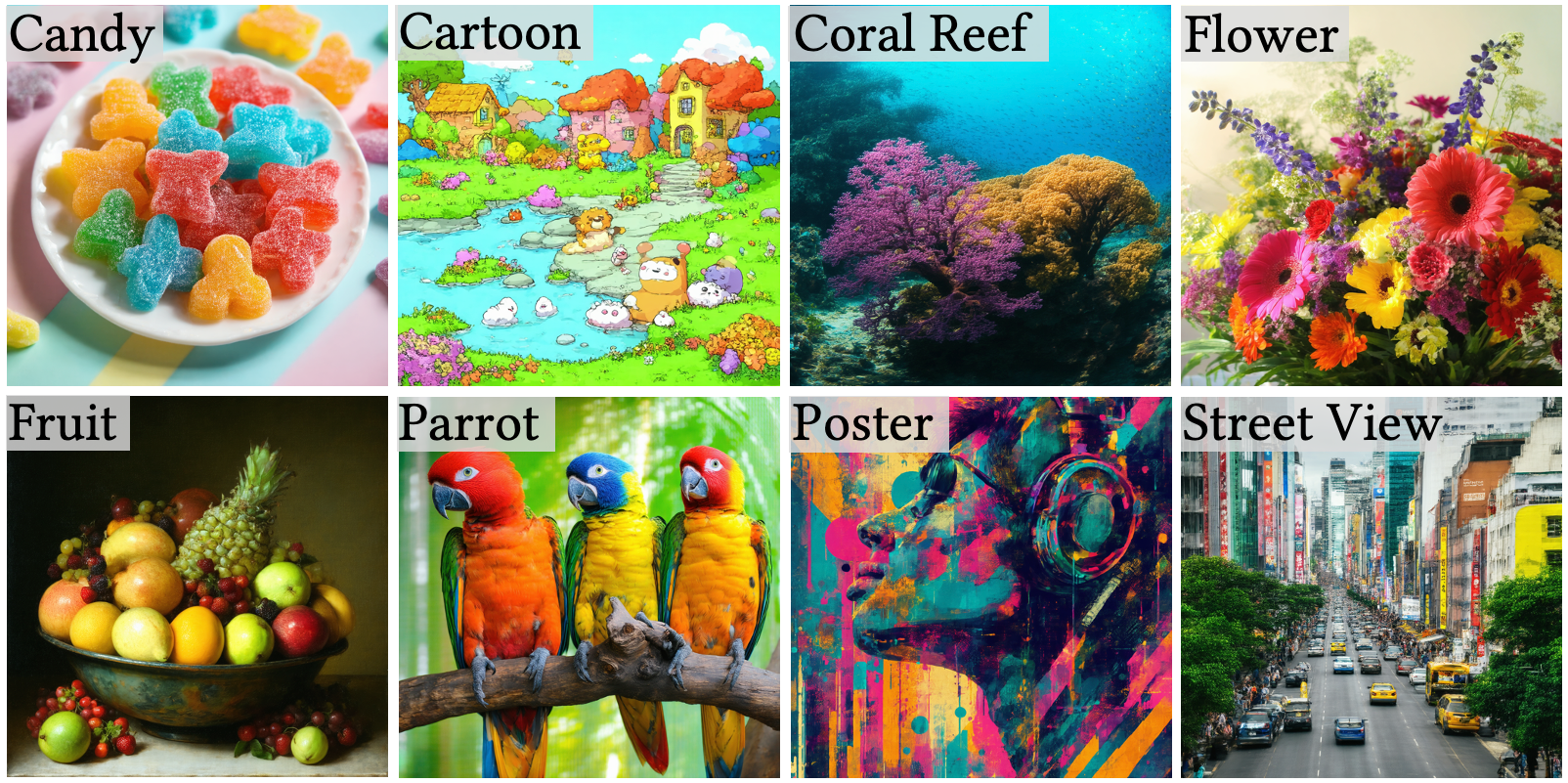}
  \vspace{-4mm}
  \caption{Example images from the eight categories in our dataset.}
  \label{fig:8example}
  \vspace{-2mm}
\end{figure}

Our work presents a systematic evaluation of color accessibility in images generated by the Stable Diffusion 3.5-large model\cite{stablediffusion}, a commonly used one in the literature. A range of four prompt strategies is considered in our study, namely \textbf{standard prompts}, \textbf{colorblind-aware prompts}, \textbf{protanopia-aware prompts} and \textbf{deuteranopia-aware prompts}. These strategies are then applied across a diverse set of content categories depicted in Figure~\ref{fig:8example}, 
%The present study does not assume that accessibility-oriented language necessarily leads to improved perceptual outcomes. Rather, it 
explicitly examining how prompt phrasing interacts with image content and color structure under simulated CVD conditions. %For each generated image, protanopia and deuteranopia are simulated using established physiologically-based models, thereby enabling controlled and repeatable evaluation.

To quantify the accessibility of a given image, we introduce a new metric that aims to capture changes in local structure, texture and edges driven by shifts in hue and saturation. To do that, we leverage perceptual color gradients \cite{Abasi2020} extracted from the image, and corresponding CVD simulations. Smaller the changes, values from our metric gets smaller, meaning a person with normal color vision and a person with a CVD would have a comparable perception of structures and textures on the image.
%gradient-based evaluation metric that measures the Mean Squared Error (MSE) between perceptual color gradient magnitude maps. The metric under discussion operates in a perceptual color space, in contrast to traditional approaches based on one-dimensional luminance gradients. It is capable of capturing changes in local structure, texture and edges driven by shifts in hue and saturation, which are critical for visual interpretation. 
%To quantify perceptual differences between normal vision and simulated CVD, we introduce a gradient-based evaluation metric that measures the Mean Squared Error (MSE) between perceptual color gradient magnitude maps. The metric under discussion operates in a perceptual color space, in contrast to traditional approaches based on one-dimensional luminance gradients. It is capable of capturing changes in local structure, texture and edges driven by shifts in hue and saturation, which are critical for visual interpretation. 
Through a series of experiments, including a synthetic verification via daltonization, our proposed metric, CVDLoss, responds consistently to accessibility-oriented color transformations% and exhibits clear category-dependent behavior
. Our results indicate that prompt-based accessibility interventions are neither reliable nor uniformly successful. This is because diffusion models are not explicitly trained to satisfy accessibility constraints. Lastly, it is also important to note that these findings position CVDLoss as a practical metric for color accessibility. %Furthermore, the study underscores the limitations of prompt engineering alone for accessibility-aware image generation.

\section{Method}
To explore color accessibility issues in generated imagery in depth, our method focuses on four core components: dataset construction, CVD simulation, and perceptual accessibility measurement and synthetic evaluation. First, since visual statistics such as color distribution and spatial structure are inherently tied to scene categories \cite{oliva2001modeling}, a diverse range of content types is necessary to systematically evaluate potential perceptual conflicts. Secondly, to achieve the transition from subjective visual inspection to objective quantification, a CVD simulator grounded in physiological principles was employed to precisely model the perceptual experiences of of protanopia and deuteranopia. Finally, addressing the insufficiency of existing luminance-based metrics in capturing hue- and saturation-driven conflicts \cite{WCAG2018}, we introduce a gradient-based perceptual metric to quantify local structural accessibility.

\subsection{Dataset Construction and Prompt Design}
% We constructed a dataset of synthetic images generated using the Stable Diffusion 3.5-large model \cite{stablediffusion}, focusing on eight visually distinct categories: \textit{building}, \textit{human face}, \textit{flower}, \textit{landscape}, \textit{fruit}, and \textit{animal}. For each category, we designed varied textual prompts:
% {\textbf{Standard prompt}}: A concise, visually descriptive phrase (e.g., “A detailed photo of a modern city building.”).
% {\textbf{Colorblind-aware prompt}}: The standard prompt augmented with “with red-green colorblind palette”.
% {\textbf{Protanopia-aware prompt}}: The standard prompt with “with protanopia-friendly palette”.
% {\textbf{Deuteranopia-aware prompt}}: The standard prompt with “with deuteranopia-friendly palette”. For each prompt type within each category, we generated a set of images (10 per prompt) using random seeds to ensure diversity.

We constructed a dataset of synthetic images generated using the Stable Diffusion 3.5-large model\cite{stablediffusion}. To capture a broad range of visual characteristics and color dependencies, we defined eight semantically and visually distinct content categories suitable for color deficiency related assessments: \textit{candy}, \textit{cartoon}, \textit{coral reef}, \textit{flower}, \textit{fruit}, \textit{parrot}, \textit{poster} and \textit{street view}. Figure~\ref{fig:8example} shows representative examples for each category in our dataset.
% \jose{If we have space (try making it), have a new figure showing an example from each category}
These categories span both color-dominant scenes (e.g., \textit{flower}, \textit{fruit}) and structure-dominant scenes (e.g., \textit{cartoon}, \textit{street view}), enabling category-level analysis of accessibility effects. For each category, we designed four types of textual prompts with increasing levels of accessibility guidance:
\begin{itemize}
\item \textbf{Standard prompt}: a concise, visually descriptive phrase (e.g., “A bowl of fruit.”).
\item \textbf{Colorblind-aware prompt}: the standard prompt augmented with the phrase “with red-green colorblind palette” (e.g., “A bowl of fruit with red-green colorblind palette.”).
\item \textbf{Protanopia-aware prompt}: the standard prompt augmented with “with protanopia-friendly palette” (e.g., “A bowl of fruit with protanopia-friendly palette.”).
\item \textbf{Deuteranopia-aware prompt}: the standard prompt augmented with “with deuteranopia-friendly palette” (e.g., “A bowl of fruit with deuteranopia-friendly palette.”).
\end{itemize}
For each prompt type and category, we generated 10 images at a resolution of $1024 \times 1024$ pixels using random seeds to encourage visual diversity. A total of 320 images were generated.
% While the use of random seeds increases sample variability, no attempt was made to ensure reproducibility across runs, as our focus lies on aggregate perceptual trends rather than exact image reconstruction.

\subsection{Simulation of Color-Vision Deficiencies}
% We obtain our CVD simulations using the implementation of\cite{Vienot1999} available in the DaltonLens library\cite{Burrus2021}. Given the space available and their prevalence in the population, we focus our study on protanopia and deuteranopia deficiencies. For each generated image, we compute protanopia and deuteranopia simulations (complete red and green blindness, respectively). 

To conduct our tests in an objective and repeatable way, we simulate how generated images would be perceived by viewers with CVD. We employ the physiologically based model proposed by\cite{Vinot1999DigitalVC}, as implemented in the DaltonLens library\cite{Burrus2021}. This model approximates the loss of one cone class by projecting colors onto the remaining cone response subspace, and has been widely adopted in both research and applied accessibility studies.
Given their prevalence in the population, we restrict our analysis to protanopia and deuteranopia CVDs. For each generated image, we run simulations at maximum severity, corresponding to complete red or green blindness (protanopia and deuteranopia, respectively). For prompts explicitly designed for a given deficiency (e.g., \textbf{protanopia-aware prompts}), only the corresponding simulation is applied. For (\textbf{standard prompts} and \textbf{colorblind-aware prompts}), both protanopia and deuteranopia simulations are computed and evaluated. The same approach would work for other types of CVDs.

\subsection{Measuring Color Accessibility}

Color conflicts due to CVD may occur between spatially adjacent regions, affecting object boundaries, textures, and fine details, or between disconnected regions with semantic importance. Detecting the latter remains a challenging open problem. In this work, we therefore focus on adjacent regions, where structural cues are encoded in image gradients.

We propose a new metric, \textbf{CVDLoss}, to quantify degradations in these structural cues under CVD simulation. For each image, we compute gradient magnitude maps for both the original image and its CVD-simulated counterpart. 
To account for perceptual 3D color differences beyond luminance, we use the color gradients from Abasi et al.\cite{Abasi2020}. For simplicity and efficiency, in our implementation we compute the HyAB color differences in the OKLab color space\cite{Ottoson2020}.
%perceptually uniform OKLab color space\cite{Ottoson2020}. Color differences between neighboring pixels are measured using the HyAB distance\cite{Abasi2020}, which combines lightness differences with chromatic differences in the $a$ and $b$ dimensions. Spatial gradients are computed using Scharr operators applied to these perceptual differences, yielding gradient magnitude maps sensitive to changes in hue, saturation, and lightness. 
%\jose{Reference to new Figure showing importance of 3D color gradients.}
%\jose{TODO: New Figure showing the synthetic red-green checkerboard pattern, where red and green have the same luminance. Next, luminance-based gradients failt o cpature any pattern on the image. Next, our gradients accurately capture the differences. I'll do it by Tuesday}
Formally, the CVDLoss between an original image $I$ and its CVD simulation $I_{\text{CVD}}$ is defined as:

\vspace{-3mm}

\begin{equation}
\text{CVDLoss}(I,I_{\text{CVD}})=
\frac{{\textstyle \sum_{p}}
\left(G(I)_{p}-G(I_{\text{CVD}})_p\right)^2}
{N \cdot \max_{p} G(I)_p^{\,2}},
\end{equation}
where $G(\cdot)$ denotes the gradient magnitude map, $p$ indexes image pixels, and $N$ is the total number of pixels.

% \jose{I want to tweak this paragraph further in combination with the caption for the Figure} 
In order to provide an intuitive understanding of what is captured by CVDLoss, Figure~\ref{fig:2mse_example} presents qualitative examples comparing normal RGB images and their protanopia simulations, along with the corresponding gradient magnitude maps. As shown, CVD can significantly modify local color gradients and edge structures in highly saturated regions, even when the global image layout appears similar. CVDLoss quantifies these perceptual–structural discrepancies by measuring differences between gradient maps, rather than relying on luminance contrast alone.

% For two color vectors $a = (L_a, A_a, B_a)$ and $b = (L_b, A_b, B_b)$, the HyAB difference is defined as:
% \begin{equation}
%    \Delta{E}_\text{HyAB}(a, b)  = \left | L_a - L_b \right | +  \sqrt{(A_a - A_b)^2+(B_a - B_b)^2}. 
% \end{equation}
% % This combines luminance and chromatic contributions, reflecting perceptual sensitivity in both channels.
% Gradient magnitude maps were then computed using Scharr operators, defined as weighted differences of HyAB values across neighboring pixels. For horizontal
% $(G_x)$ and vertical $(G_y)$ directions:
% \begin{equation}
%    G_x(i, j) = w_0 \cdot \Delta{E(z_1, z_7)} + w_2 \cdot \Delta{E(z_2, z_8)} + w_0 \cdot \Delta{E(z_3, z_9)},
% \end{equation}
% \begin{equation}
%    G_y(i, j) = w_0 \cdot \Delta{E(z_1, z_3)} + w_2 \cdot \Delta{E(z_4, z_6)} + w_0 \cdot \Delta{E(z_7, z_9)},
% \end{equation}
% where $z_1...z_9$ denote the $3\times3$ neighborhood of a pixel, and $w_0 = 47.0$, $w_1 = 162.0$ are Scharr operators.

% The overall gradient magnitude at pixel $(i, j)$ is:
% \begin{equation}
%    G(i, j)  =  \sqrt{G_x(i, j)^2+G_y(i, j)^2}. 
% \end{equation}
% % Finally, the Gradient-MSE score was defined as the mean squared error between the gradient magnitude maps of the original and CVD-simulated images.
% Finally, the Gradient-MSE score was defined as the mean squared error between the gradient magnitude maps of the original $(G_{orig})$ and CVD-simulated $(G_{cvd})$ images.
% \begin{equation}
%    \text{Gradient-MSE}  =  \frac{1}{N}  {\textstyle \sum_{i=1}^{N}}({G_{orig}(i)-G_{cvd}(i))^2}. 
% \end{equation}

\subsection{Synthetic Verification via Daltonization Interpolation}
To further validate the effectiveness of the proposed CVDLoss metric, we conduct a synthetic verification experiment based on image daltonization. Daltonization optimizes the original colors so viewers with CVDs can perceive the modified content better. 

For our purposes, this means a daltonized image should suffer a smaller loss of gradients than the original ones, resulting in a smaller CVDLoss value.
For each category, we apply full daltonization\cite{daltonizelib} to images generated with the standard prompt, targeting protanopia and deuteranopia. We then measure the change in CVDLoss as
$\log_{10}(\text{CVDLoss}_{\text{daltonized}}) - \log_{10}(\text{CVDLoss}_{\text{original}})$,
where CVDLoss compares each image to its corresponding CVD simulation.

\begin{figure}[h]
  \centering
  \includegraphics[width=\linewidth]{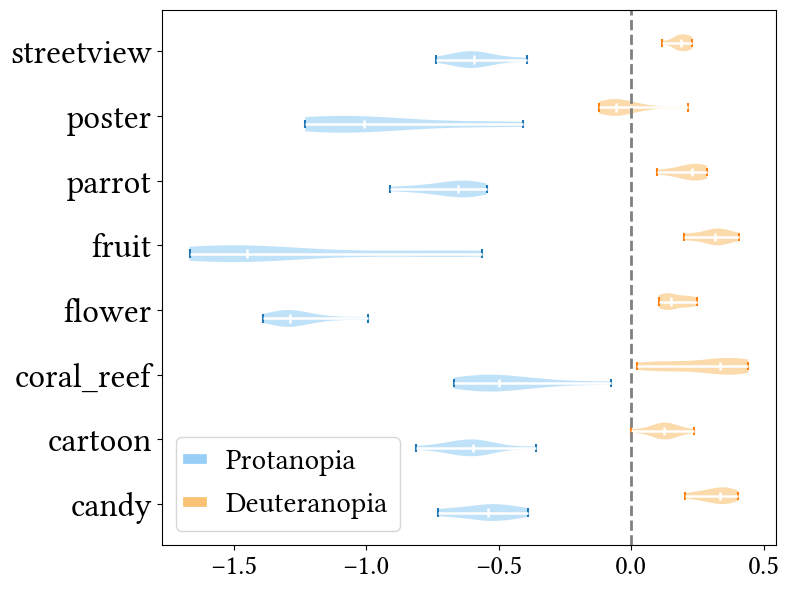}
  \vspace{-4mm}
  \caption{Distribution of differences in $\log_{10}
  (\text{CVDLoss})$ between original images and their daltonized images for protanopia (blue) and deuteranopia (orange) across categories, capturing relative changes in structural discrepancies induced by daltonization. The vertical zero line indicates no change in CVDLoss after daltonization; positive values indicate increased distortion, so negative values are preferred. 
  % \jose{Do we want to add that results for Deuteranopia are unexpected, but still consistent? We treat daltonization as a black box that may not optimize for the kind of loss we are measuring, and so resulting strategies for optimizing for deuteranopia may differ from protanopia.}
  }
  % \jose{Shouldn't this plot appear flipped vertically? Based on all we discussed for Figure 3, daltonized results should preserve gradients better under CVD simulations, so their Gradient-MSE should be lower than the input images, and so appear below the 0 line, right?  }
  \vspace{-2mm}
  \label{fig:violin1}
\end{figure}

Figure~\ref{fig:violin1} shows the distribution of this change across eight categories. A consistent reduction in CVDLoss after daltonization is observed for protanopia across most categories, suggesting that daltonized images tend to preserve local gradient structure more faithfully under protanopia simulation. In contrast, deuteranopia often exhibits positive shifts, particularly in color-rich categories. This difference is expected, as daltonization methods are treated here as a black box and are not explicitly designed to minimize gradient-based losses. For deuteranopia, higher post-daltonization contrast may increase relative gradient discrepancies, even when absolute contrast is preserved or enhanced.

\FloatBarrier
\FloatBarrier
% To further validate the effectiveness of the proposed Gradient-MSE metric, we conduct a synthetic verification experiment based on image daltonization. Daltonization solutions optimize the original colors so viewers with CVD can perceive the modified content better. 
% For our purposes, this means a daltonized image should suffer a smaller loss of gradients than the original ones, resulting in a smaller CVDLoss value.
% For each category, we apply full daltonization\cite{daltonizelib} to images generated with the standard prompt, targeting protanopia and deuteranopia. We then measure the change in CVDLoss as
% $\log_{10}(\text{CVDLoss}_{\text{daltonized}}) - \log_{10}(\text{CVDLoss}_{\text{original}})$,
% where CVDLoss compares each image to its corresponding CVD simulation.

% Figure~\ref{fig:violin1} shows the distribution of this change across eight categories. A consistent reduction in CVDLoss after daltonization is observed for protanopia across most categories, suggesting that daltonized images tend to preserve local gradient structure more faithfully under protanopia simulation. In contrast, deuteranopia often exhibits positive shifts, particularly in color-rich categories. This difference is expected, as daltonization methods are treated here as a black box and are not explicitly designed to minimize gradient-based losses. For deuteranopia, higher post-daltonization contrast may increase relative gradient discrepancies, even when absolute contrast is preserved or enhanced.

Importantly, the proposed metric captures both behaviors in a stable and interpretable manner across categories and deficiencies, rather than producing noise-like variations. This confirms that CVDLoss responds meaningfully to accessibility-oriented color transformations, supporting its validity as a perceptual diagnostic metric for color accessibility evaluation.

% In most categories, Gradient-MSE increases after daltonization, particularly for color-rich content such as \textit{candy}, \textit{flower}, \textit{fruit}, and \textit{parrot}. This phenomenon aligns with expectations: daltonization introduces substantial hue remapping, resulting in increased variations in local color-gradient structures within the CVD simulation. Conversely, categories with simpler or more constrained color structures, such as \textit{cartoon} and \textit{poster}, exhibit smaller or mixed changes. \jose{Update this discussion}

% Crucially, the metric responds consistently to the daltonization process across categories and deficiencies, rather than producing noise-like variations. This confirms that Gradient-MSE is sensitive to perceptually meaningful color transformations and correlates with known accessibility-driven color modifications, supporting its validity as a metric for evaluating color accessibility.

% Importantly, the metric responds consistently to the daltonization process across categories and deficiencies, producing systematic and interpretable changes rather than noise-like variations. This behavior confirms that Gradient-MSE is sensitive to perceptually meaningful color transformations and correlates with known accessibility-oriented color remapping strategies. As such, this synthetic verification supports the validity of Gradient-MSE as an evaluation metric for color accessibility, independently of prompt design or subjective user input. \jose{This paragraph feels too long. Make it shorter to save space}.

\section{Experimental Results}

\begin{figure}
  \centering
  \includegraphics[width=\linewidth]{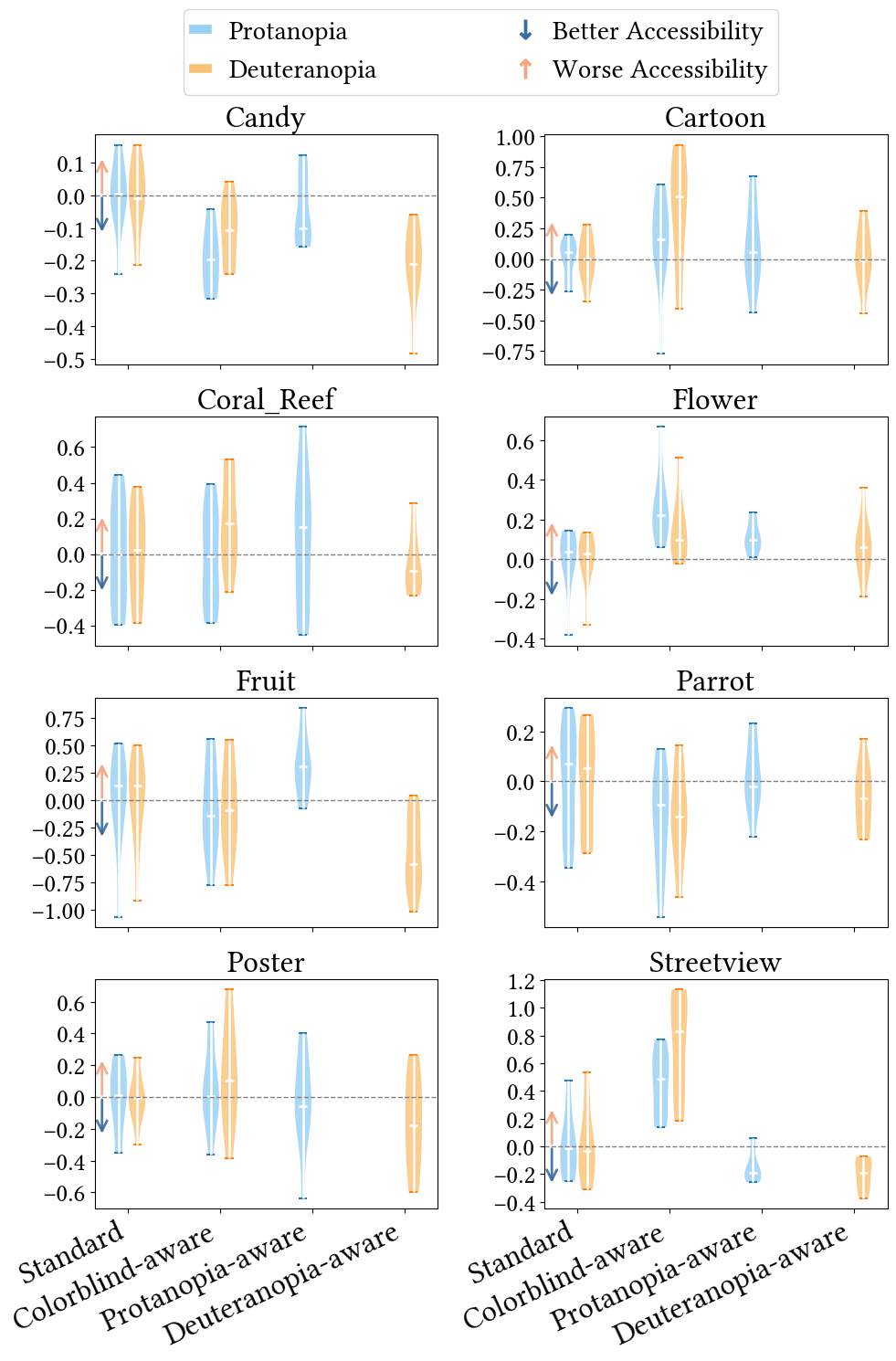}
  \vspace{-4mm}
  \caption{Normalized CVDLoss values for each category, prompt, and CVD in log scale. Each subplot corresponds to a category.
  % blue and orange respectively represent Protanopia and Deuteranopia deficiencies. 
  Values are normalized by subtracting the mean of $\log_{10}(\text{CVDLoss})$ for the standard prompt (horizontal dashed line at zero). Arrows indicate the trends of accessibility improvements or reductions at the perceptual-structural level.}
  \label{fig:violin2}
  \vspace{-2mm}
\end{figure}

As illustrated in Figure ~\ref{fig:violin2}, the normalized CVDLoss distributions are summarized across eight content categories, four prompt types and two forms of CVD. Negative values indicate reduced perceptual–structural discrepancy under CVD simulation relative to the standard prompt, while positive values indicate increased discrepancy.

Compared to Figure~\ref{fig:violin1}, the impact of accessibility-oriented prompts is category- and deficiency-dependent, with no universally positive trend. It is evident that categories characterized by strong hues, such as \textit{candy}, and \textit{flower}, demonstrate the most significant variability in normalized CVDLoss. This finding suggests that prompt-induced color changes have a substantial impact on perceptual–structural consistency under the context of CVD simulation. However, these effects differ in direction: while \textit{candy} generally benefits from accessibility-oriented prompts, \textit{flower} consistently shows increased CVDLoss, suggesting that color reinterpretation often disrupts rather than preserves local structure in this category.

\textit{Cartoon}, \textit{poster} and \textit{street view}, exhibit a similar response pattern across prompt strategies. All three categories show a clear instability under the \textbf{colorblind-aware prompt}, characterized by a widened distribution and increased CVDLoss, indicating heightened perceptual disruption. Deficiency-specific prompts partially mitigate this effect, but to different extents. While \textit{cartoon} and \textit{poster} only show modest reductions with protanopia- and deuteranopia-aware prompts, \textit{street view} largely return to distributions comparable to the \textbf{standard prompt} with a substantially stronger decrease in CVDLoss. This suggests that deficiency-specific prompting can reduce perceptual disruption, but its effectiveness strongly depends on scene composition and the interaction between color structure and spatial layout.

% \FloatBarrier
% \FloatBarrier

Other categories, including \textit{coral reef} and \textit{parrot}, exhibit only modest shifts across prompts, indicating limited sensitivity to color-based prompt interventions. \textit{Fruit} further highlights the asymmetry between deficiencies, with protanopia- and deuteranopia-aware prompts producing opposite effects, reinforcing the need to evaluate CVD separately.

% \begin{figure}
%   \centering
%   \includegraphics[width=\linewidth]{violin2_v3_newmetric.png}
%   \vspace{-4mm}
%   \caption{Normalized CVDLoss values for each category, prompt, and CVD in log scale. Each subplot corresponds to a category.
%   % blue and orange respectively represent Protanopia and Deuteranopia deficiencies. 
%   Values are normalized by subtracting the mean of $\log_{10}(\text{CVDLoss})$ for the standard prompt (horizontal dashed line at zero). Arrows indicate the trends of accessibility improvements or reductions at the perceptual-structural level.}
%   \label{fig:violin2}
%   \vspace{-4mm}
% \end{figure}

\section{Conclusion \& Discussion}
Collectively, these findings demonstrate that diffusion models are not explicitly trained to understand accessibility constraints. Consequently, the interpretation of accessibility-related language in prompts is not consistent, resulting in outcomes that are both unpredictable and, occasionally, unfavorable. Rather than ensuring enhanced accessibility, prompt-based interventions have the potential to introduce instability, particularly in color-dominant scenes. In this context, CVDLoss could function as an effective diagnostic metric, capturing nuanced perceptual–structural changes across content and deficiencies, and helping to identify where current generative models fail to produce consistent, accessible outputs.

Our analysis is limited to a single diffusion-based model and a constrained set of prompt formulations. However, without explicit accessibility supervision during training, we do not expect prompt engineering alone to yield consistently reliable improvements across models or prompts. The proposed CVDLoss metric focuses on losses of local perceptual structure and texture, and does not address accessibility issues arising from non-adjacent or semantically related regions, which remains an open problem. In addition, our synthetic verification relies on a single daltonization method treated as a black box, leading to differing behaviors across deficiencies. Analyzing more method, evaluating alternative daltonization pipelines and validating the metric through user studies with CVD individuals are identified as significant directions for future research.

\bibliographystyle{eg-alpha-doi} 
\bibliography{egbibsample}

\end{document}